\documentstyle[11pt,paspconf,epsf]{article}

\def\plotone#1{\centering \leavevmode
\epsfxsize=\columnwidth \epsfbox{#1}}

\def\plot08#1{\centering \leavevmode
\epsfxsize=0.8\columnwidth \epsfbox{#1}}

\begin{document}

\title{Radio Frequency Interference}

\author{R. D. Ekers and J. F. Bell}
\affil{ATNF CSIRO, PO Box 76 Epping NSW 1710, Sydney Australia; rekers@atnf.csiro.au~~~jbell@atnf.csiro.au}

\begin{abstract}
We describe the nature of the interference challenges facing radio astronomy
in the next decade. These challenges will not be solved by regulation only,
negotiation and mitigation will become vital. There is no silver bullet for
mitigating against interference. A successful mitigation approach is most
likely to be a hierarchical or progressive approach throughout the telescope
and signal conditioning and processing systems. We summarise some of the
approaches, including adaptive systems.

\end{abstract}

\keywords{radio astronomy, interference, mitigation, OECD, adaptive systems,
communications}

\section{The RFI  Challenge and Spectrum Management}

\begin{figure}[htbp]
\begin{center}
\plotone{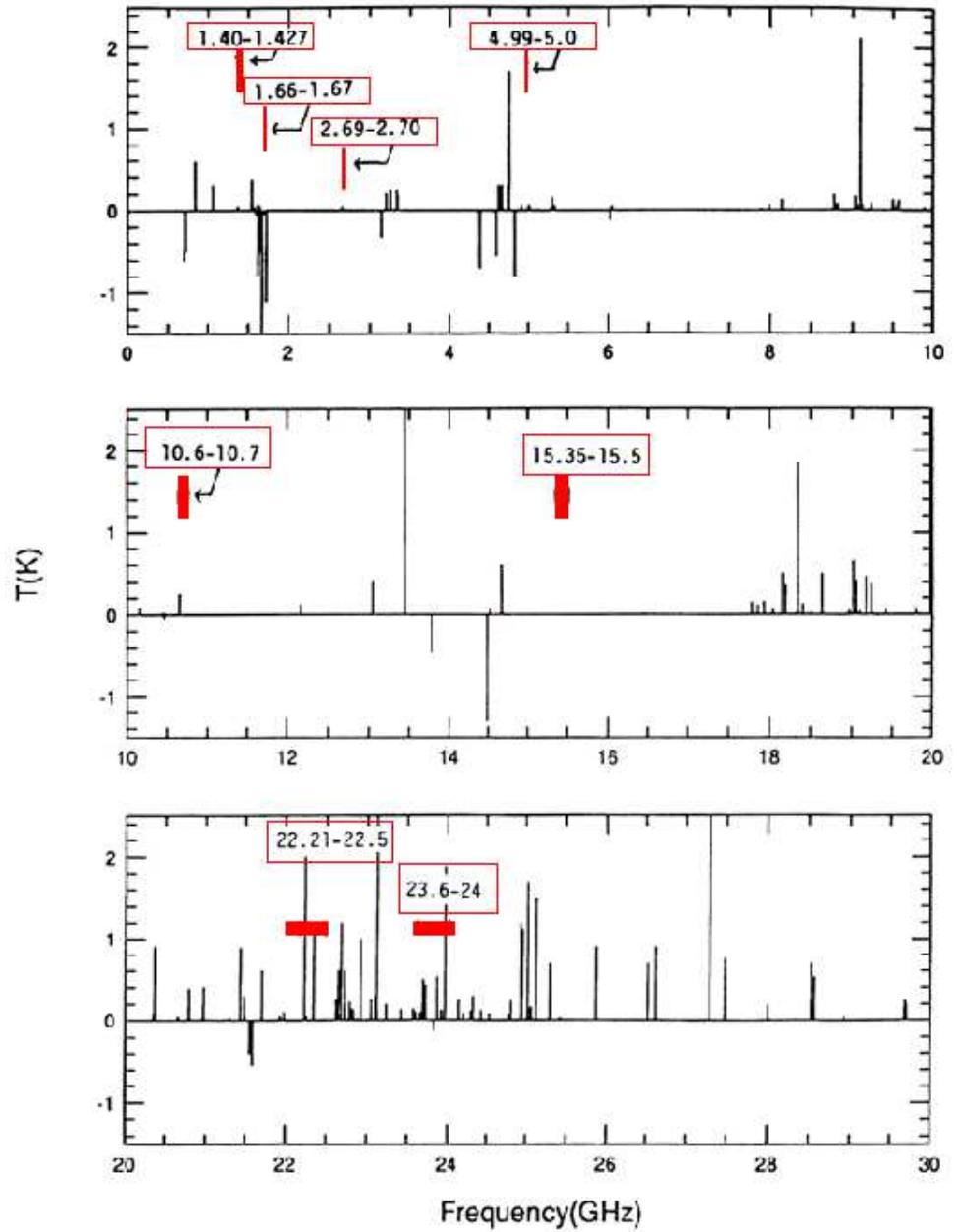}
\caption{Spectral lines (at zero redshift) are indicated in absorption or
emission over the 0--30 GHz band. The boxes indicate the bands allocated for
passive radio astronomy uses. Figure from Morimoto 1993.}
\label{alloc}
\end{center}
\end{figure}

If future telescopes like the SKA are developed with sensitivities up to 100
times greater than present sensitivities, it is quite likely that current
regulations will not provide the necessary protection against
interference. There is a range of experiments (eg redshifted hydrogen or
molecular lines) which require use of arbitrary parts of the spectrum, but
only at a few locations, and at particular times, suggesting that a very
flexible approach may be beneficial. Other experiments require very large
bandwidths, in order to achieve enough sensitivity. As shown in Figure
\ref{alloc}, presently only 1-2\% of the spectrum in the metre and
centimetre bands is reserved for passive uses, such as radio astronomy
(Morimoto 1993). In the millimetre band, much larger pieces of the spectrum
are available for passive use, but the existing allocations are not
necessarily at the most useful frequencies. Current regulations alone will
be inadequate, we need technology as well as regulation.  We cannot (and do
not want to) impede the telecommunications revolution, but we can try to
minimise its impact on passive users of the radio spectrum and maximise the
benefits of technological advances. Further information on many of the
topics discussed below is available on http://www.atnf.csiro.au/SKA/intmit/.

\section{Classes of Interference}

It is important to be clear of what we mean when we talk about
interference. Radio astronomers make passive use of many parts of the
spectrum legally allocated to communication and other services.  As a
result, many of the unwanted signals are entirely legal and legitimate. We
will adopt the working definition that interference is any unwanted signal
entering the receiving system.

Interfering signals vary a great deal in their source and nature. This
naturally leads to different mitigation approaches. Local sources of
interference include things internal to telescope instruments, networking
for IT systems, and general and special purpose digital processors in the
observatory. Interference compliance testing, shielding, separate power
circuits, minimising nearby equipment are key steps that need to be taken to
minimise this kind of interference.

External interference may arise from fixed or moving sources. Not all
methods of mitigation apply to both: in fact methods that work well for
fixed sources, may not work at all for moving sources, due to problems like
side lobe rumble. Interference may be naturally occurring or human
generated.  Examples of naturally occurring interference include: the
ground, sun, other bright radio sources, and lightening.  Human
generated interference may come from broadcast services (eg TV, radio),
voice and data communications (eg mobile telephones, two-way radio, wireless
IT networks), navigation systems (eg GPS, GLONASS), radar, remote sensing,
military systems, electric fences, car ignitions, and domestic appliances
(eg microwave ovens) (Goris 1998).

The vast majority of these operate legally with in their allocated bands,
regulated by national authorities and the ITU (International
Telecommunications Union). However there are sources of interference, such
as the Iridium mobile communications systems, whose signals leak into bands
protected for passive use. In this case, these interfering signals are
$10^{11}$ times stronger than the signal from the early universe.  In the
case of Australia, there is a single communications authority for whole
country and therefore for the whole continent. As a result there is a single
database containing information on the frequency, strength, location, etc of
every licensed transmitter (Sarkissian 2000). A key point therefore, is that
the modulation schemes and other characteristics of the vast majority of
these signal are known. Their effect on radio telescopes is not only
predictable, but can be modelled and used to excise the unwanted signals.

\begin{figure}[htbp]
\begin{center}
\plotone{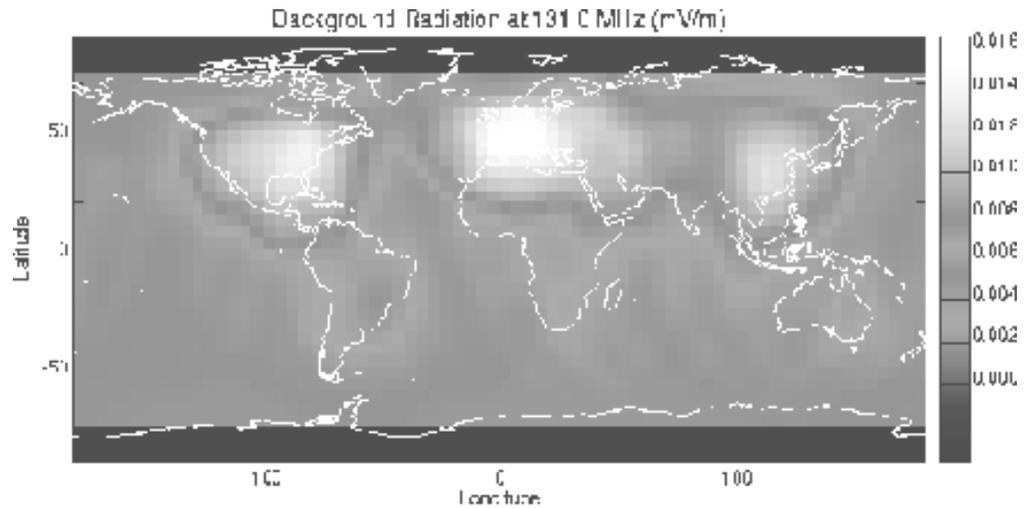}
\caption{Forte satellite: 131MHz, RF background data for calibration of
lightening monitors provides a map of signals emanating from the Earths
surface, many of which are human generated. Acquired by the Los Alamos
National Laboratory FORTE spacecraft, Principal Investigator
A. R. Jacobson. Data courtesy of R. J. Strangeway, UCLA.}
\label{forte}
\end{center}
\end{figure}

Radio astronomy could deal with most terrestrial interfering signals, by
moving to a remote location, where the density and strength of unwanted
signals is greatly reduced. As shown in Figure \ref{forte}, this is getting
more and more difficult, but there are still some possibilities.  However
with the increasing number of space borne telecom and other communications
systems in low (and mid) Earth orbits, a new class of interference
mitigation challenges are arising - radio astronomy can run, but it cannot
hide !  The are several new aspects introduced into the interference
mitigation problem by this and they include: rapid motion of the transmitter
on satellites, more strong transmitters in dish side lobes and possibly in
primary beam, and different spectrum management challenges, because no place
on Earth is free from interference from the sky.

\section{RFI fundamentals}
Undesired interfering signals and astronomy signals can differ (be
orthogonal) in a range of parameters, including: frequency, time, position,
polarisation, distance, coding, positivity, and multi path. It is extremely
rare that interfering and astronomy signals do not possess some level of
orthogonality in this $\geq 8$ dimensional parameter space. We therefore
need to develop sufficiently flexible signal processing systems to take
advantage of the orthogonality and separate the signals. This is of course
very similar to the kinds of problems faced by mobile communication
services, which are being addressed with smart antennas and software radio
technologies. Examples in radio astronomy to date include the use of the
time or frequency phase space or even better both, as in pulsar studies
wherer the time/frequency dispersion relation can be used, and the
requirement that signals are positive in very low frequency studies. Antenna
arrays could take advantage of the position and distance (curvature of
wavefront) phase space. Human generated interference is normally polarised,
so unpolarised astronomical signals can be observed by measuring the
unpolarised component $(I - (U^2 + Q^2 + V^2)^{1/2})$ (R. Fisher, private
communication.

\subsection{Mitigation Strategies and Issues}

There is no silver bullet for detecting weak astronomical signals in the
presence of strong undesired naturally occurring or human generated
signals. Spectral bands allocated for passive use provide a vital window,
which cannot be achieved in any other way. It is important to characterise
the RFI so that the number, strength, band width, duty cycle, spatial and
frequency distributions, and modulation and coding schemes can all be used
to advantage in modelling and mitigating RFI. Doing this at low frequencies
gives greater sensitivity due to the effects of harmonic content and ease of
propagation. In order to do this, the telescope and instruments must be
calibrated to provide the best possible characterisation of interfering and
astronomy signals.  A/D converters must be fast enough to give sufficient
bandwidth, with a sufficient number of bits so that both strong and weak
signals are well sampled.  There are a range of techniques that can make
passive use of other bands possible and in general these need to be used in
a progressive or hierarchical way.

\noindent
$\bullet~~~${\bf  Remove at source} is obviously best, but that is often not possible,\\
$\bullet~~~${\bf  Regulation} providing radio quiet frequencies or regions,\\
$\bullet~~~${\bf  Negotiation} with owners users can lead to win-win solutions,
for example replacing nearby radio links with underground fibres, removes
interference and improves voice and data connectivity for users,\\
$\bullet~~~${\bf  Avoid interference} by choosing appropriate locations with
terrain screening or radio quiet zones,\\
$\bullet~~~${\bf  Move to another frequency},\\
$\bullet~~~${\bf  Screening} to prevent signals entering the primary elements
of receivers,\\
$\bullet~~~${\bf  Far Side lobes} of primary and secondary elements must be both
minimised and well characterised,\\
$\bullet~~~${\bf  Minimise coherent signals} through out the array and thereby
allow the natural rejection of the array to deal with the incoherent
signals,\\
$\bullet~~~${\bf  Front end filtering}, using for example high temperature
super conducting filters with high Q to reject strong signals in narrow
bands, before they cause saturation effects.\\
$\bullet~~~${\bf  High dynamic range linear receivers} to allow appropriate
detection of both astronomy (signals below the noise) and very strong
interfering signals,\\ 
$\bullet~~~${\bf  Notch filters} (analog, digital or photonic) to excise
bad spectral regions,\\
$\bullet~~~${\bf  Clip} samples from data streams to mitigate burst type
interference,\\
$\bullet~~~${\bf  Decoding} to remove signals with complex modulation and
multiplexing schemes. Blanking of period
or time dependent signals is a very successful but simple case of this
more general approach,\\
$\bullet~~~${\bf  Cancellation} of undesired signals, before correlation using
fixed and adaptive signal processing (harris, 2000),\\
$\bullet~~~${\bf  Post correlation cancellation} of undesired signals,
taking advantage of phase closure techniques (Sault 2000)\\
$\bullet~~~${\bf  Parametric techniques} allow the possibility of taking
advantage of known interference characteristics to excise it (Ellingson,
2000),\\ 
$\bullet~~~${\bf  Adaptive beam forming} to steer one or more nulls onto
interfering 
sources. This is equivalent to cancellation, but it provides a
way of taking advantage of the spatial orthogonality of astronomy and
interfering signals,\\
$\bullet~~~${\bf  Use of Robust statistics} in data processing to minimise the
effects of outliers.

\subsection{Which signal processing regime: traditional analog, digital, photonic ?}

In most applications of signal processing, there is a strong trend towards
the use of digital techniques, as well as photonic techniques. The
fundamental reason for this is that digital and photonic devices have cost
curves which are evolving much more rapidly than traditional analog
systems. In addition to that, they open up new techniques and offer
substantial reduction in computational effort in many cases. The inherent
immunity of photonic approaches to radio interference also creates
functional advantages (Minasian 2000). Astronomers are joining these trends
for exactly the same reasons. The jury is still out on what the appropriate
balance or mixture of these techniques will be.

\section{Adaptive EMI rejection}

Adaptive rejection algorithms can be either constrained or
unconstrained. Constrained algorithms generally incorporate either a model
or a copy of the desired or interfering signal, which is used control the
adaption. For example it may be constrained so that only those signals with
a certain coding or chip sequence are removed (Ellingson, 2000). Most
astronomy signals are expected to be pure noise, so one could envisage a
constraint that rejects all non noise like signals. In the case of
unconstrained adaption, some algorithms (predictive adaptive algorithms)
simply assume that the interference is much stronger than the signal and
just use previous data samples to predict the following data samples for
cancellation (harris, 2000). If that assumption cannot be made, another
approach is to block the desired signal and let the unconstrained algorithm
work on the remaining signals. This is often done using a blocking matrix,
which can be thought of as an operator that applies a set of complex weights
which block certain signals, while passing everything else. Advantages of
this approach are that it can deal with multiple interfering signals which
are changing in time and space, without affecting the signal to noise of the
desired signal.

A key ingredient of constrained adaptive algorithms is a reference channel
that maximises interference to noise ratio for the ensemble of
interferers. One way of achieving this is using additional omni-directional
antennas or arrays which at least matches the gain of the side lobe response
of the main array.

\subsection{Adaptive Interference Cancelling}

Of all the approaches listed above, the nulling or cancellation systems (may
be adaptive or predictive) are the most likely to permit the observation of
weak astronomy signals that are coincident in frequency or space with
undesired signals. There is an important space-time duality with cancelling
algorithms. Any algorithm that works in the time domain can also be applied
in the space domain.

\begin{figure}[htbp]
%\vspace{1.75in}
\plot08{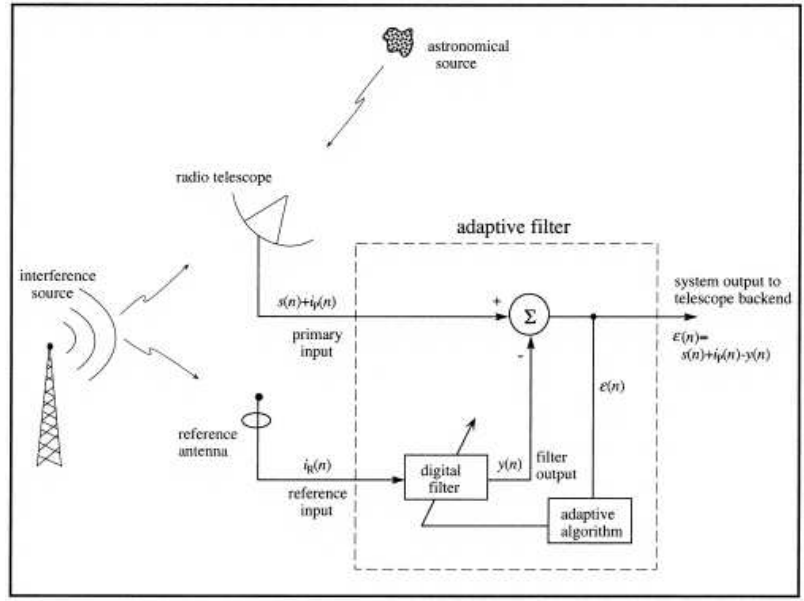}
\caption{A Conceptual View of Adaptive Interference Cancelling. From
Barnbaum \& Bradley (1998).} \label{fig-2}
\end{figure}

These techniques have been used extensively in communications, sonar, radar,
medicine and others (Widrow \& Stearns 1985, Haykin 1995). Radio astronomers
have not kept pace with these developments and in this case need to infuse
rather than diffuse technology in this area. A prototype time based
cancellation system developed at NRAO (shown in Figure \ref{fig-2}) has
demonstrated 70dB of rejection on the lab bench and 30dB of rejection on
real signals when attached to the 140 foot at Green Bank (Barnbaum \&
Bradley 1998). Adaptive nulling systems are being prototyped by NFRA in the
Netherlands (van Ardenne, these proceedings). Combined space-time approaches
have been used to cancel interference in GPS receivers (Trinkle, 2000).
However, in all cases, the application in the presence of real radio
astronomy signals is yet to be demonstrated and their effects on the weak
astronomy signals needs to be quantified. A good prospect for doing this in
the near future is recording baseband data from existing telescopes,
containing both interfering and astronomy signals and simulating the
receiver system in software (Bell et al. 1999). A number of algorithms can
then be implemented is software and assessed relative to each other.

Beam forming and adaptive nulling as are not necessarily being done
sequentially, but rather in parallel.  While there are some sequential
schemes (genetic algorithms for example), most approaches simultaneously
solve for the coefficients that give the desired beams and nulls.  One can
think of this as an optimisation problem.  For example, in the minimum
variance beam former, the ``goal'' is to minimise output power, and the
``constraint'' is maintain constant gain in a certain direction.  The goal
forces the nulls onto all interfering signals not coming from the direction
of the astronomical source, and the constraint protects the beam gain
(Ellingson, 2000).  Of course, it is only protected for one direction, so
there is still shape distortion. In general for an N element array, you can
form up to N-1 nulls. However if more control over the main beam shape is
required, one may use other beam formers, which form a smaller number of
nulls, and use more degrees of freedom to control the main beam shape.

A physical interpretation of why you can form nulls without wrecking the
beam might go as follows: Phased arrays with many elements (not equally
spaced), have lots of nulls, and they are all over the sky once you get a
reasonable distance from the main beam.  Imagine changing the coefficients a
little to get the closest one on to an interferer.  Very little variation in
the coefficients is required.  Since the difference is so small, the main
beam hardly notices.  The other nulls will shift around, of course, because
they are sensitive to small changes in the coefficients.  Close to the main
beam, the nulls are further apart, so you need a bigger variation in the
coefficients to nudge the closest null into place - hence the increased
distortion in the main beam in this case. It may be necessary to record the
weights applied to generate the nulls, so that the beam shape changes can be
calibrated out later (Cram, 2000).

\section{Real time v post correlation}

Real time systems permit full recovery of temporal information in the signal
required.  Real time systems have been well studied and numerous examples
can be found in other fields such as: radar, sonar, communications, defence
anti jam, speech processing, and medicine.  For example, there are existing
systems which are capable of nulling up to 7 simultaneous moving jammers. In
radio astronomy we operate in a totally different regime in which the
astronomical signals are weak and noise like. We only wish to measure the
time averaged statistical properties of the signals. For example in
aperture synthesis the time averaged coherence between two antennas. Since
we don't have to recover the signal modulation, radio astronomy does not have
to use the real time algorithms developed for communications and radar. In
such post correlation systems (Sault 2000) the information is only contained
in the statistical properties of the signal which may vary slowly in time,
frequency, space or direction. Both the signal of interest and the
interference obey phase and amplitude closure relations. This results in an
over determined set of equations which form a closed set which can be used
to self calibrate the array for both the source and the interferer.

We conjecture that: {\bf {\it ``post correlation processing of time averaged
signals can achieve the same RFI rejection as in real-time algorithms and
that self calibration (phase closure) techniques provide powerful additional
constraints''}}.

\end{document}